\documentclass[prd,preprint,tightenlines,floatfix,showpacs,showkeys,date,preprintnumbers,nofootinbib,eqsecnum]{revtex4}
\usepackage{graphicx,longtable,array}
  \usepackage{bm}
   \usepackage{amsmath}
    \usepackage{amssymb}
     \usepackage{pifont}
      \input epsf.sty
\newcommand{\be}{\begin{equation}}\newcommand{\ee}{\end{equation}}%
\newcommand{\bd}{\begin{displaymath}}\newcommand{\ed}{\end{displaymath}}
\newcommand{\bit}{\begin{itemize}}                                        
 \newcommand{\eit}{\end{itemize}}                                         
\newcommand{\ben}{\begin{enumerate}}                                      
 \newcommand{\een}{\end{enumerate}}                                       
\newcommand{\baa}{\begin{array}{lll}}                                     
 \newcommand{\eaa}{\end{array}}                                           
\newcommand{\ba}{\begin{eqnarray}}                                        
 \newcommand{\ea}{\end{eqnarray}}                                         
\newcommand{\la}{\label}                                                  
\newcommand{\Ds}{\displaystyle}                                           
\newcommand{\gev}[1]{\relax\ifmmode{\text{GeV}^{#1}}                      
                     \else{GeV$^{#1}${ }}\fi}                             
\def\MSbar{\relax\ifmmode\overline                                        
            {\rm MS}\else{$\overline{\rm MS}${ }}\fi}                     
\def\as{\relax\ifmmode \alpha_s\else{$ \alpha_s${ }}\fi}                  
\def\abar{\relax\ifmmode{\bar{a}}\else{$\bar{a}${ }}\fi}                  
  \def\ie{\hbox{\it i.e.}{ }}                 
   \def\eg{\hbox{\it e.g.}{ }}                  
 
\def\1{\hbox{{1}\kern-.25em\hbox{l}}}
\newlength{\tabcolf} 
\newlength{\tabcols} 
\newlength{\tabcolt} 

\begin{document}
\date{\today}

\title{ ERBL and DGLAP kernels for transversity distributions.\\
{\small Two-loop calculations in covariant gauge}}

\author{S.~V.~Mikhailov}%
 \email{mikhs@theor.jinr.ru}
\author{A.~A.~Vladimirov}
\email{avlad@theor.jinr.ru}
\affiliation{%
  Bogoliubov Laboratory of Theoretical Physics,
  JINR, 141980, Moscow Region, Dubna, Russia}%

\vspace {10mm}

\begin{abstract}
The results of a two-loop calculation in the Feynman gauge
of both the DGLAP and the ERBL evolution kernels for 
transversely polarized distributions are presented.
The structure of these evolution kernels is discussed
in detail.
In addition, the effect of the two-loop evolution on the 
distribution amplitude of a twist-2 transversely polarized 
meson is explored.
\end{abstract}
\vspace {2mm}

\maketitle
\newpage
\section{I\lowercase{ntroduction}}
Evolution kernels are the main ingredients of the well-known evolution
equations for the parton distribution in DIS processes  \cite{L75}
and for the parton distribution amplitudes \cite{BL80} in hard exclusive
reactions. 
These equations describe the dependence of the parton
distributions on the
renormalization scale $\mu^2$. 
Previously, two-loop calculations
were performed for the unpolarized forward DGLAP evolution
kernel $P(z)$ in \cite{L75,CFP80,FLK81} and, what is more cumbersome, 
for the nonforward ERBL kernel
$V(x,y)$ in \cite{BL80,DR84,MR85} that was challenging and complicated
technical tasks.
Here, 
we present the results of a direct calculation of evolution kernels
for the transversity distributions in next-to-leading-order (NLO) 
performed in the \MSbar scheme.
These calculations are carried out in the
Feynman gauge within a single
mold for both the forward kernels and the nonforward ones, i.e.,
\begin{eqnarray}
P^T(x) &=& a_s\ P^T_0(x)~~~+a_s^2\ P^T_1(x)~~~+ \ldots \ , \\
V^T(x,y) &=& a_s\ V^T_0(x,y)+a_s^2\ V^T_1(x,y)+ \ldots \ ,
\end{eqnarray}
where $a_s= \alpha_s(\mu^2)/(4\pi)$.

Note that the kernel $P^T_1$ was first obtained in \cite{Vogelsang98} within
a light-cone gauge calculation and shortly thereafter the corresponding anomalous
dimensions $\gamma^T_1(n)$ were presented in \cite{KM1997,HKK1997}.
The kernel $V^T_1$ was reconstructed in \cite{BFM2000} on the basis of the knowledge
of the structure of symmetry-breaking terms for the kernel, 
which first appeared at the two-loop level.
For the reader's convenience, let us explain these issues in more detail.
Those terms of $V^T_1$ that are
responsible for the conformal-symmetry breaking can be
fixed and expressed via some special convolutions of the known \cite{MR86ev,Mul95,BFM2000}
one-loop kernel elements.
At the same time, the remaining part or, in other words,
the symmetrical part (in terms of a conformal-group representation) of this kernel
can eventually be
restored from a certain part of the forward kernel $P^T_1$ \cite{BFM2000}.
This possibility of ``guessing'' will not be pursued here.

The calculation of $V^T$ or $P^T$
can be performed following the standard procedure
to find the renormalization-group generators in the \MSbar scheme.
Expressions for them in terms of the renormalization
constant $Z_{\Gamma}$ for every diagram $\Gamma$ in
(see, \eg, \cite{V80} and  \cite{MR85}) are given by
\ba
Z_{\Gamma}=1 - \hat{K}R'(\Gamma),
~~~~V~(\ P\ )=-a_s\partial_{a_s}\left(Z^{(1)}_{\Gamma}\right) =
a_s\partial_{a_s}\left(\hat{K}_1 R' (\Gamma) \right)
\stackrel{\rm NLO}{\longrightarrow} 2\hat{K}_1 R' (\Gamma).
\la{Z}
\ea
Here, (i) $R'$ is the incomplete BPHZ $R$-operation; $D = 4-2\varepsilon$
is the space-time dimension,
(ii)  $\hat{K}$ separates out poles in $\varepsilon$, whereas
$\hat{K}_1$ picks out
a simple pole,  and (iii) $Z^{(1)}$ is the
coefficient of the simple pole in the expansion of $Z_{\Gamma}$.
To introduce an appropriate notation for the analysis of the two-loop results,
let us start with the
leading order $V^T_0~(P^T_0)$ results obtained in a covariant
$\bm{\xi}$-gauge\footnote{
The gauge
parameter $\bm{\xi}$ is defined via the gluon propagator in lowest-order 
perturbation theory which reads $\Ds i D^{a b}_{\mu \nu}(k^2) = \frac{-i
\delta^{ab}}{k^2+i\epsilon} \left(g_{\mu \nu} -
\bm{\xi}\frac{k_{\mu}k_{\mu}}{k^2} \right)$
},
\begin{eqnarray}
 \label{eq:P0}
P^{T}_0(x)&=&C_F~ \left[p_0(x)_+ - \delta(1-x)\right]\ , \\
\label{eq:V0}
V^{T}_0(x,y)&=&C_F~ \left[2F^T(x,y)_+ + (x\to \bar{x},~y\to \bar{y}) \right]- \delta(y-x)
\end{eqnarray}
Here, $p_0(x) \equiv \Ds \frac{4x}{1-x}$;  $\Ds F^T(x,y) \equiv \frac{x}{y}\frac{1}{y-x}$;
 $\bar{x}=1-x,~\bar{y}=1-y$;
symbol $(\ldots)_+$ denotes different distributions like
$p(x)_+ = p(x) - \delta(1-x)\int^1_0 p(z)\ dz$ and $V(x,y)_+ = V(x,y)- \delta(y-x)\int^1_0 V(z,y)\ dz$.
The diagrammatic expansion of the kernels is presented in the Table below,
where $\bm{\xi}$-dependent terms appear in the partial diagrams $a, c$
canceling out each other in the complete results in Eqs.\ (\ref{eq:P0}--\ref{eq:V0}),
 as expected.
\vspace*{-3mm}
\begin{table*}[h]
\caption{Diagrammatic expansion of the one-loop kernels with 
$MC$ denoting the mirror--conjugated diagrams \label{tab:1-loop}}
 \hspace*{-6mm}
 \begin{tabular}{||c|c|c||}\hline
 \begin{picture}(60,60)
\put(-9,-45){\includegraphics[scale=0.5]{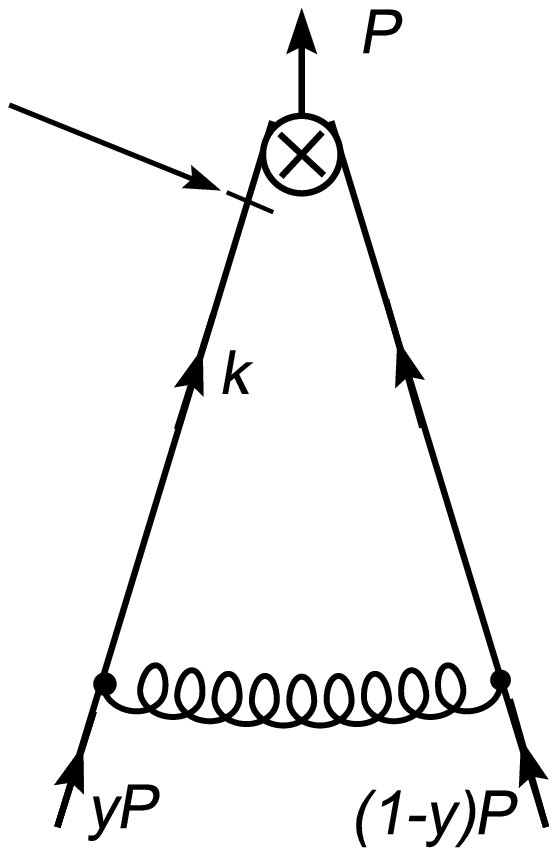}}
\put(-25,68){\makebox{$\bm{\delta(x-\frac{nk}{nP})}$}}
\end{picture}&
\raisebox{-1.5cm}{\includegraphics[scale=0.5]{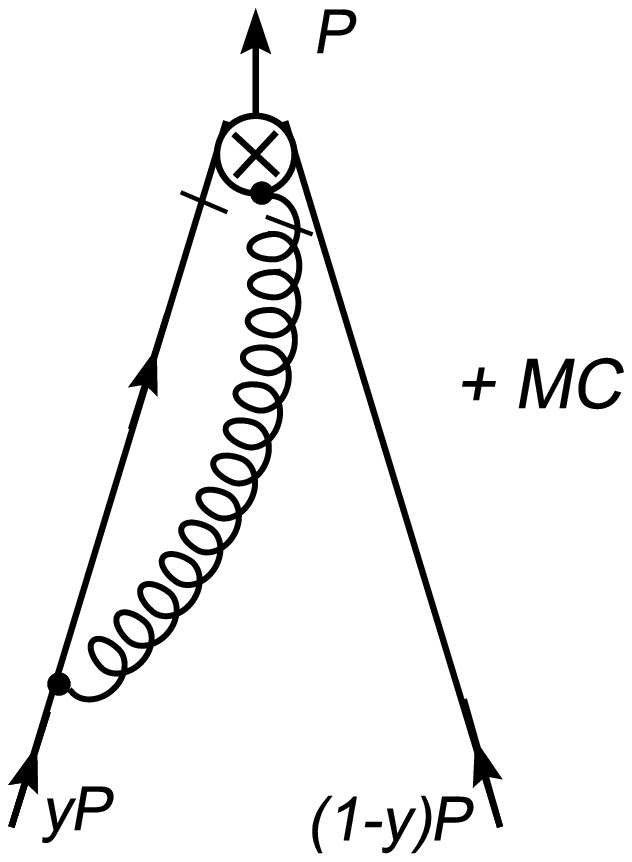}}&
\raisebox{-1.5cm}{\includegraphics[scale=0.5]{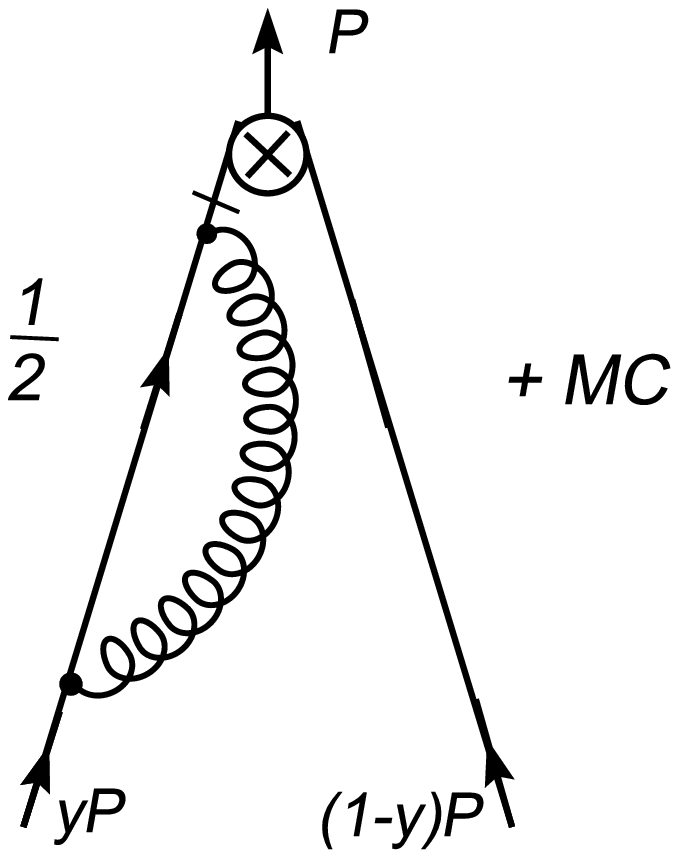}} \\

\begin{tabular}{ccc}
$P_a(x)$&=&$- C_F \bm{\xi}\delta(1-x)$ \\ ~ \\
$V_a(x,y)$&=&$- C_F
\bm{\xi}\delta(y-x)$
\end{tabular} &

\begin{tabular}{ccc}
  & &                                              \\
$P_b(x)$&=&$\Ds  \ C_F \Big(p_0\Big)_+ \equiv
C_F\Big(\frac{4x}{1-x}\Big)_+ $\\~\\
$V_b(x,y)$&=&$\Ds  \ C_F \Big(\mathcal{C}\theta(y>x)\ 2 F^T\Big)_+$\\~\\
\end{tabular} &

\begin{tabular}{ccc}
$P_c(x)$&=&$- C_F(1-\bm{\xi}) \delta(1-x) $\\~\\
$ V_c(x,y)$&=&$-C_F(1-\bm{\xi}) \delta(y-x)  $
\end{tabular}
\\
\hline
\end{tabular}
\end{table*}
\vspace{-3mm}
\noindent
The slash on the line of each of these diagrams denotes the delta function $\delta(x-nk/nP)$, 
where $k$ is the momentum on this line, while $n$ is a light-cone vector ($n^2=0$).
These diagrammatic calculation rules can be traced to the momentum representation
of the composite operator
$\bar{\psi}(0)\sigma_{\mu \nu}\psi(\lambda n)$, denoted here by $\otimes$,
and were elaborated in detail in~\cite{MR85}.
The abbreviation $MC$ in the figures denotes the mirror--conjugate diagrams, 
while the symbol $\mathcal{C}$ denotes the corresponding mirror conjugation
of arguments, $\mathcal{C} \theta(y>x)f(x,y)\equiv $ $\theta(y>x)f(x,y)+ \theta(y<x)f(\bar{x},\bar{y})$.
The local current, corresponding to the operator $\otimes$, is not conserved
and, therefore, there is no ``plus'' prescription imposed on Eqs.\ (\ref{eq:P0}), (\ref{eq:V0}).
Therefore, the separate $\delta$-functions survive.

Noting that the product $\left(y \bar{y}\right) V^{T}_0(x,y)$ is symmetric under the exchange
$x\leftrightarrow y$,
one realizes that the corresponding anomalous dimension matrix can be diagonalized
in the Gegenbauer basis $\{\psi_n(x)=(x\bar{x})C^{3/2}_n(2x-1)\}$.
The deeper reason for this is that conformal symmetry survives at the LO level \cite{Mak81}.
On the other hand, at NLO the conformal symmetry does not hold true (in the \MSbar scheme) owing to
renormalization effects.
These generate specific terms in $V^T_1$
that break this $x \leftrightarrow y$ symmetry as well as the diagonalization property
mentioned above.
For brevity, the corresponding terms will be referred to as ``nondiagonal (diagonal)'' ones.

In the next section, the contributions to $P^T_1$ and $V^T_1$ for
each of the 2-loop diagrams will be demonstrated explicitly.
In Sec.\ \ref{sec:structure},
 we analyze the structure of both calculated kernels
which are in accord with the expected manifestation of these
symmetry breaking terms.
Finally, we confirm the results for $P^T_1$, calculated in
\cite{Vogelsang98}, as well as the result for $V^T_1$ found in \cite{BFM2000}.
The kernel $V^T_1$ provides the key ingredient, necessary for any complete
NLO analysis of exclusive processes involving
transversely polarized vector mesons via a QCD evolution of their distribution amplitudes.
For an illustration of this NLO evolution,
we analyze in Sec.\ \ref{sec:evolution} how it affects
the transversely polarized $\rho$-meson distribution amplitude
(DA) at the characteristic scale $\mu^2_B$ applicable to the $B$-meson
semi-leptonic decay \cite{BM2000}.
Our main findings are summarized in Sec.\ \ref{sec:conclusion} together with our conclusions.

\section{D\lowercase{iagram-by-diagram presentation for} $\bm{P^T_1}$ \lowercase{and} $\bm{V^T_1}$}
 \label{sec:diagrams}
Here, we present the diagram-by-diagram results of the calculation of the
DGLAP, $P^T_1$, and ERBL, $V^T_1$, kernels at the two-loop level for  $\bm{\xi}=0$.
In all there are 19 diagrams in the list below where we also
display the diagrams with a zero contribution.
The full list of them can be found in \cite{FLK81}.
Note that superscripts {\large $\star$} mark the obtained new result for each
diagram.
The results for the other diagrams can be restored from those obeying 
the DGLAP \cite{FLK81} or ERBL \cite{MR85} evolution kernels.
Diagrams {\large f$^\star$} and {\large h} with gluon-loop insertions include 
also the corresponding ghost loops.
Let us remark that there are
only four basic scalar topologies of integrals, the latter being presented 
\footnote{see also corrections to these results in Appendix B in \cite{OVel03}}
in \cite{MR85}.

 \begin{longtable}{||p{3cm}|c||}\hline
\Large{d}\raisebox{-1.5cm}{\includegraphics[scale=0.33]{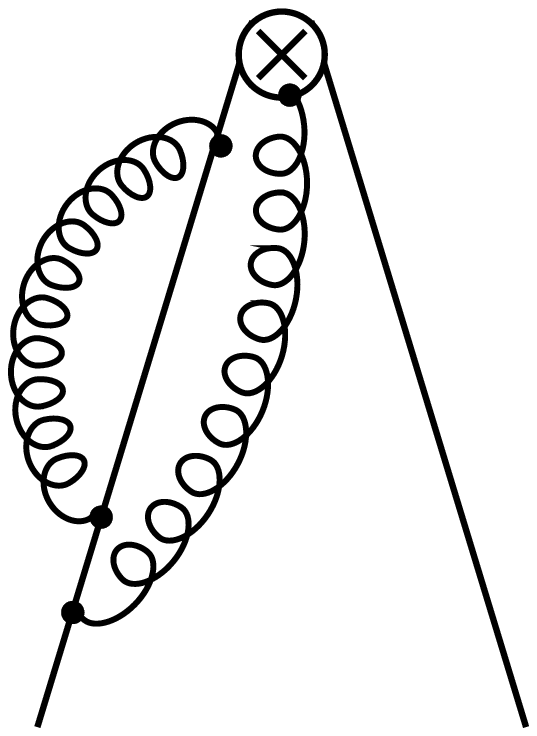}}

 & \begin{tabular}{cc p{10cm}}\\
 P(x) &=& $\Ds -C_F^2\Big[p_0\Big(1+\ln \bar x\Big)\Big]_+$
        \\\\
         \hline
          \\
 V(x,y)&=&$\Ds -2C_F^2~\Big[\mathcal{C}\theta(y>x)F^T\Big(1+\ln\Big(1-\frac{x}{y}\Big)\Big)\Big]_+$
\end{tabular}

\\ \hline
\Large{e}$^\star$\raisebox{-1.5cm}{\includegraphics[scale=0.33]{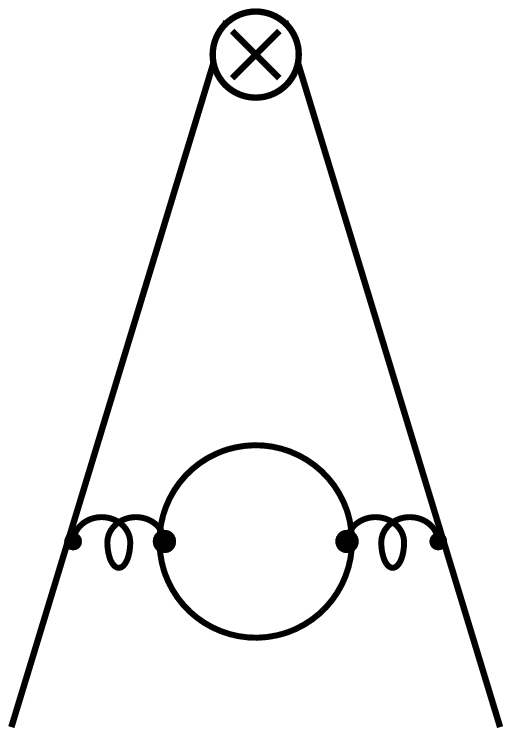}}

& \begin{tabular}{cc p{10cm}}\\
 P(x) &=& $\Ds C_FT_rN_f\frac{8}{9}\delta(1-x)$
        \\\\
        \hline
        \\
 V(x,y)&=&$\Ds C_FT_rN_f\frac{8}{9}\delta(y-x)$
\end{tabular}
\\ \hline
    \Large{f}$^\star$ \raisebox{-1.5cm}{\includegraphics[scale=0.33]{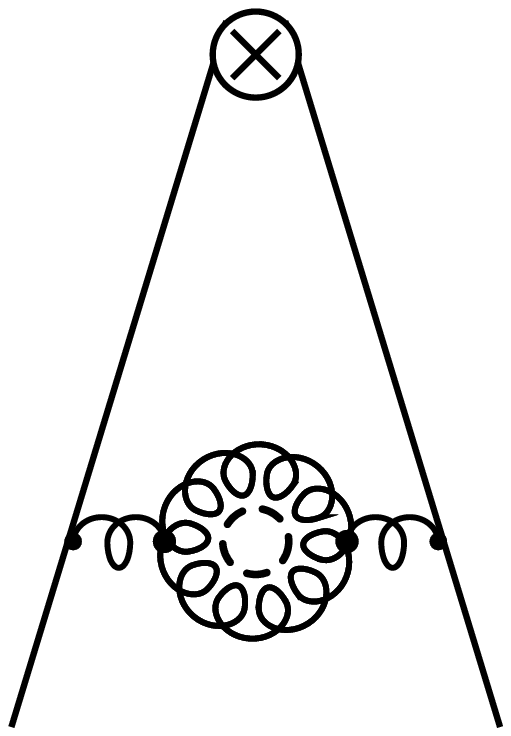}}

& \begin{tabular}{cc p{10cm}}
\\
 P(x) &=& $\Ds -C_FC_A\frac{16}{9}\delta(1-x)$
        \\\\
         \hline
          \\
 V(x,y)&=&$\Ds -C_FC_A\frac{16}{9}\delta(y-x)$
\end{tabular}

\\ \hline
\Large{g}\raisebox{-1.5cm}{\includegraphics[scale=0.33]{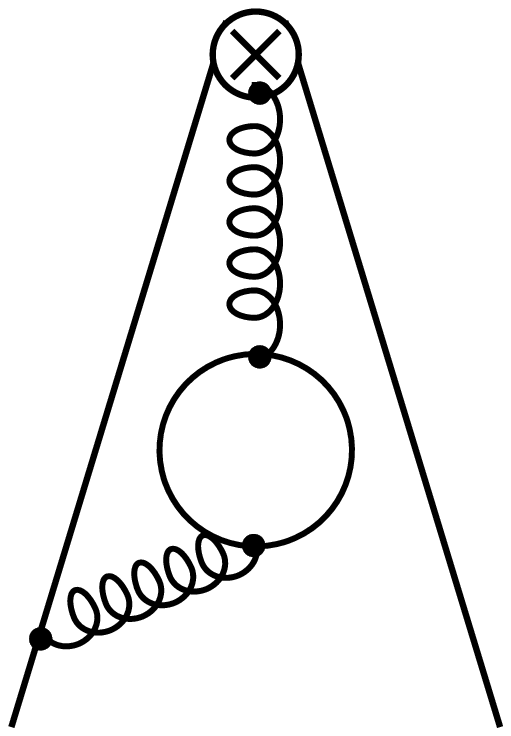}}

& \begin{tabular}{cc p{10cm}}
\\
 P(x) &=& $\Ds -C_FT_rN_f\Big[p_0\Big(\frac{20}{9}+\frac{4}{3}\ln x\Big)\Big]_+$
        \\\\
        \hline
        \\
 V(x,y)&=&$\Ds -2C_FT_rN_f~\Big[\mathcal{C}\theta(y>x)F^T\Big(\frac{20}{9}+\frac{4}{3}\ln \frac{x}{y}\Big)\Big]_+$
\end{tabular}

\\ \hline
\Large{h}\raisebox{-1.5cm}{\includegraphics[scale=0.33]{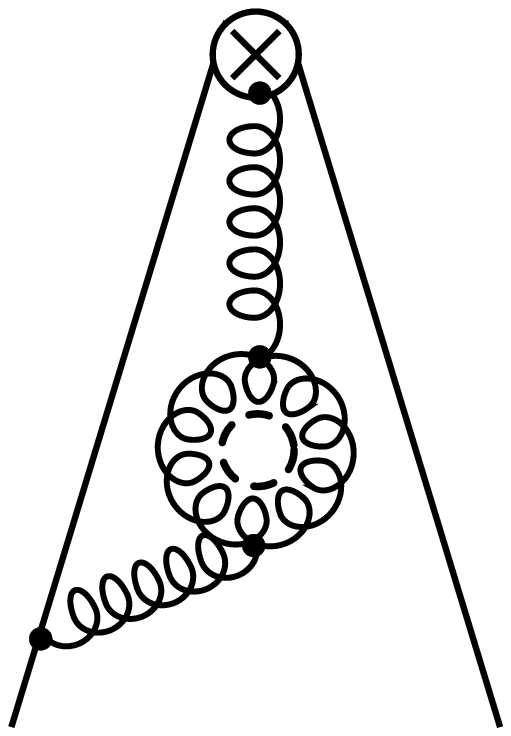}}

& \begin{tabular}{cc p{10cm}}
\\
 P(x) &=& $\Ds C_FC_A\Big[p_0\Big(\frac{31}{9}+\frac{5}{3}\ln x\Big)\Big]_+$
        \\\\
        \hline
        \\
 V(x,y)&=&$\Ds 2C_FC_A~\Big[\mathcal{C}\theta(y>x)F^T\Big(\frac{31}{9}+\frac{5}{3}\ln \frac{x}{y}\Big)\Big]_+$
\end{tabular}

\\ \hline
\Large{i}$^\star$ \raisebox{-1.5cm}{\includegraphics[scale=0.33]{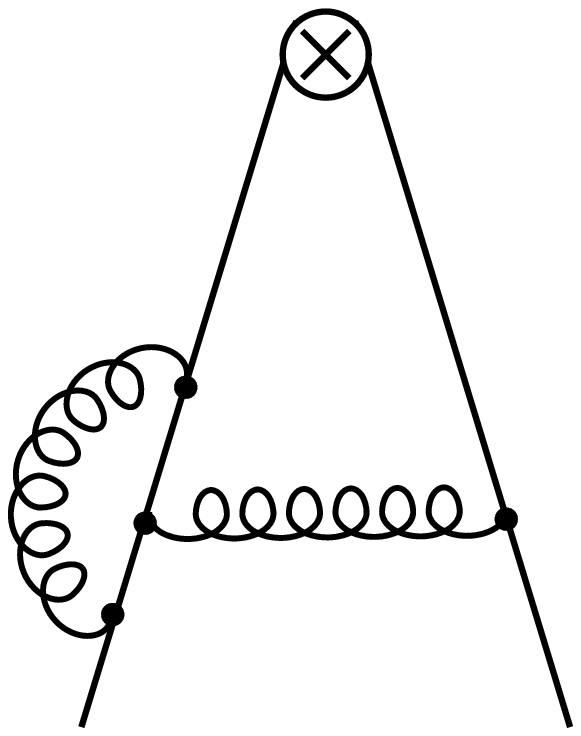}}

& \begin{tabular}{cc p{10cm}}
\\
 P(x) &=& $\Ds -4C_F\Big(C_F-\frac{C_A}{2}\Big)~\ln\bar x$
        \\\\
        \hline
         \\
 V(x,y)&=&$\Ds -2C_F\Big(C_F-\frac{C_A}{2}\Big)\mathcal{C}\theta(y>x)\frac{1}{y}\ln\big(1-\frac{x}{y}\big)$
\end{tabular}

\\ \hline
\Large{k}$^\star$ \raisebox{-1.5cm}{\includegraphics[scale=0.33]{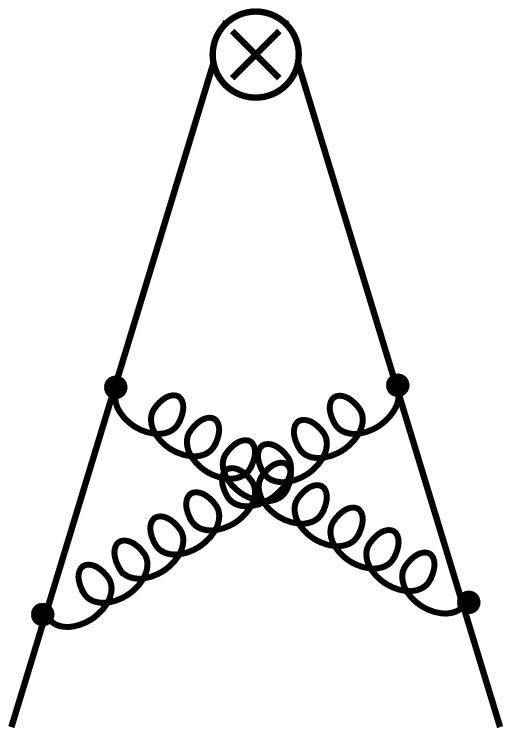}}

& \begin{tabular}{cc p{10cm}}
\\
 P(x) &=& $\Ds 4C_F\Big(C_F-\frac{C_A}{2}\Big)~\Big(\bar x+\eta \bar x\Big)$
        \\\\
         \hline
          \\
 V(x,y)&=&$\Ds 4C_F\Big(C_F-\frac{C_A}{2}\Big)\mathcal{C}\Big[\theta(y>x)\frac{x}{y}+
 \theta(y>\bar x)\frac{\bar x}{y}\Big]$
\end{tabular}
\\ \hline
\Large{l}\raisebox{-1.5cm}{\includegraphics[scale=0.33]{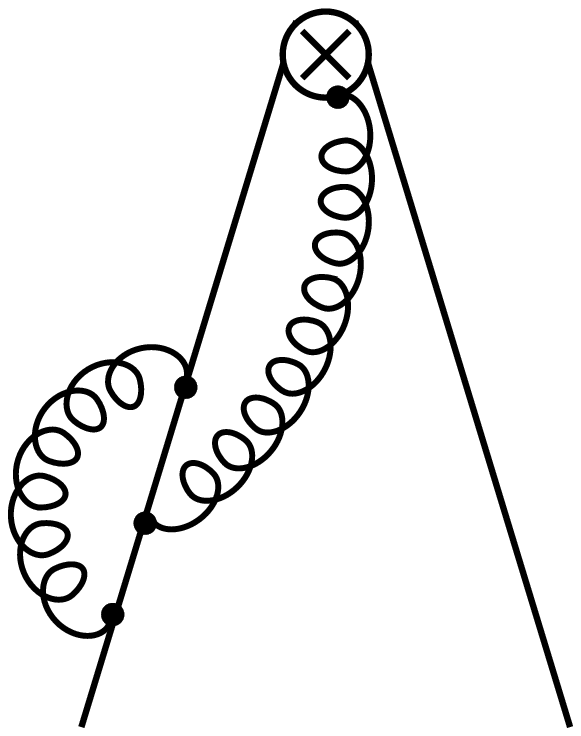}}

& \begin{tabular}{cc p{10cm}}
\\
 P(x) &=& $\Ds C_F\Big(C_F-\frac{C_A}{2}\Big)~\Big[p_0\Big(1-3\ln x-\ln^2x+\ln\bar x\Big) + 12 \ln\bar x\Big]_+$
        \\\\
         \hline
          \\
 V(x,y)&=&$\Ds 2C_F\Big(C_F-\frac{C_A}{2}\Big)
 \Bigg\{\mathcal{C}\theta(y>x)\Big[F^T\bigg(1-3\ln\frac{x}{y}-\ln^2\frac{x}{y}
 $ \\ &&$\Ds+\ln\Big(1-\frac{x}{y}\Big)\bigg)
  +\frac{3}{y}\ln\Big(1-\frac{x}{y}\Big)\Big]\Bigg\}_+$

\end{tabular}
\\ \hline
\Large{m}\raisebox{-1.5cm}{\includegraphics[scale=0.33]{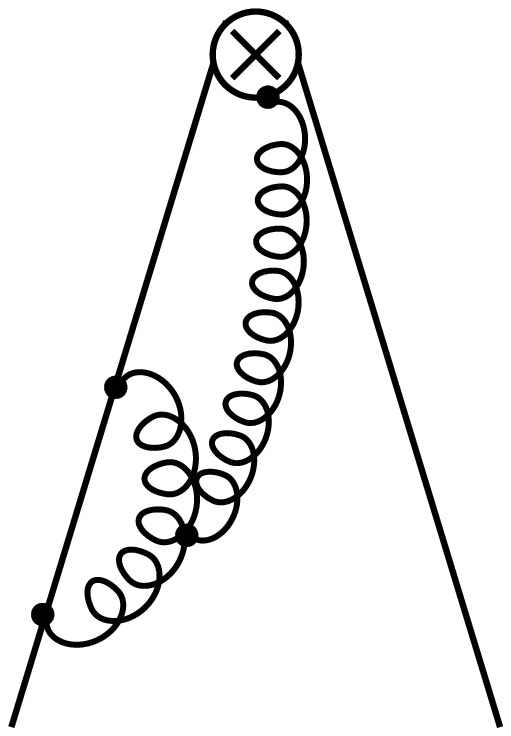}}

& \begin{tabular}{cc p{10cm}}
\\
 P(x) &=& $\Ds C_FC_A~\Big[p_0\Big(\frac{3}{2}-\frac{1}{2}\ln x-\frac{1}{4}\ln^2\bar x\Big) + \ln^2\bar x\Big]_+$
        \\\\
        \hline
         \\
 V(x,y)&=&$\Ds C_FC_A\Bigg\{\mathcal{C}\theta(y>x)\Big[F^T\bigg(3-\ln\frac{x}{y}-\frac{1}{2}\ln^2\Big(1-\frac{x}{y}\Big)\bigg)
$ \\
&&$\Ds+\frac{1}{2y}\ln^2\Big(1-\frac{x}{y}\Big)\Big]\Bigg\}_+$\\~
\end{tabular}

\\ \hline
\Large{n}$^\star$ \raisebox{-1.5cm}{\includegraphics[scale=0.33]{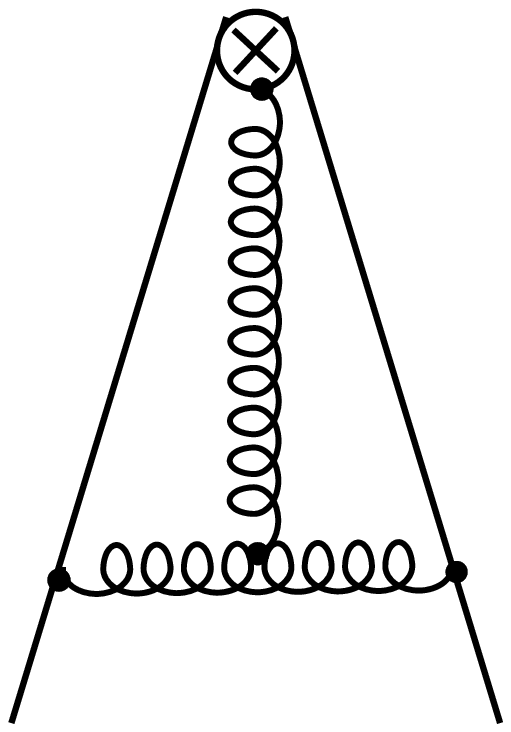}}

& \begin{tabular}{cc p{10cm}}
\\
 P(x) &=& $\Ds C_FC_A\Bigg[\ln^2\bar{x} +4\ln\bar{x}$\\
&&~~~~~~~~~~$\Ds +p_0\Big(\frac{1}{4}\ln^2 x-\text{Li}_2(1-x) -
\ln\bar{x} \ln x\Big)\Bigg]_+$
\\ \\
\hline
\\

 V(x,y)&=&$\Ds C_FC_A\Bigg\{\mathcal{C}\theta(y>x)\Big[
 \frac{1}{2y}\ln^2\Big(1-\frac{x}{y}\Big)+
\frac{2}{y}\ln\Big(1-\frac{x}{y}\Big)+ \frac{2}{y \bar{y}}\ln
x\ln\bar{x}$ \\ &&$ \Ds
 + F^T\Big[\frac{1}{2}\ln^2\frac{x}{y}-
2\text{Li}_2\Big(1-\frac{x}{y}\Big) \Big]+2\bar
F^T\ln\frac{x}{y}\ln\Big(1-\frac{x}{y}\Big)\Big] \Bigg\}_+$
\end{tabular}
\\ \hline
\Large{o}$^\star$ \raisebox{-1.5cm}{\includegraphics[scale=0.33]{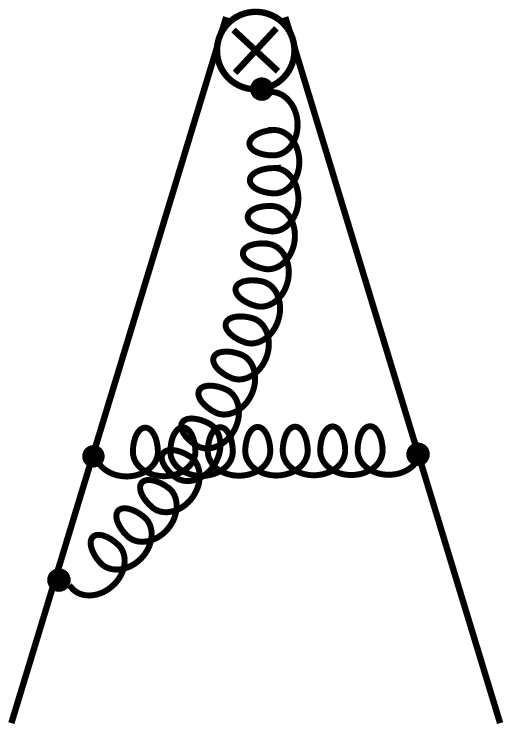}}

& \begin{tabular}{cc p{10cm}}
\\
 P(x) &=& $\Ds C_F\Big(C_F-\frac{C_A}{2}\Big)\Bigg[p_0\Big(4 \text{Li}_2(1-x)-\ln ^2x\Big)+8 \ln\bar{x}$
 \\ &&$\Ds +\eta~
 2p_0\bigg(\text{Li}_2\Big(\frac{|x|}{1+|x|}\Big)-\text{Li}_2\Big(\frac{1}{1+|x|}\Big)+\frac{1}{2}\ln^2|x|$ \\ &&$\Ds-\ln|x|\ln(1+|x|)\bigg)\Bigg]$
\\ \\
\hline
\\

 V(x,y)&=&$\Ds 2C_F\Big(C_F-\frac{C_A}{2}\Big) \Bigg\{\mathcal{C}\theta(y>x)
\Big[\frac{2}{y}\ln\Big(1-\frac{x}{y}\Big) +\frac{2}{y\bar y}\ln
x\ln \bar x$
\\ &&$\Ds -\frac{2}{y\bar
y}\ln\frac{x}{y}\ln\Big(1-\frac{x}{y}\Big)
+F^T\bigg(4\text{Li}_2\Big(1-\frac{x}{y}\Big)+\ln^2\frac{x}{y}\bigg)\Big]+G^T(x,y)\Bigg\}$

\end{tabular}

\\ \hline
\Large{q}\raisebox{-1.5cm}{\includegraphics[scale=0.33]{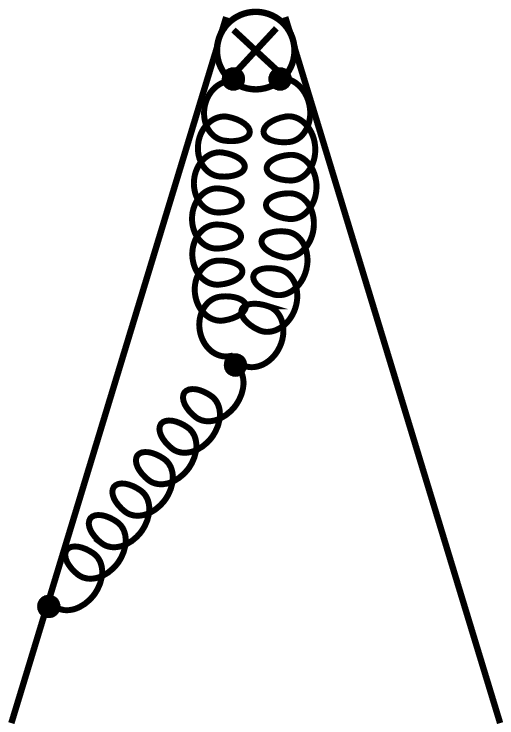}}

& \begin{tabular}{cc p{10cm}}
 \\
 P(x) &=& $\Ds
C_FC_A\Big[p_0\Big(1+\ln x +S(x)+\frac{1}{4}\ln^2
x-\frac{1}{2}\ln\bar x+\frac{1}{4}\ln^2\bar x\Big)$
 \\ && $\Ds -4\ln\bar
x-2\ln^2\bar x \Big]_+$
\\\\
 \hline
  \\
V(x,y)&=&$\Ds C_FC_A\Bigg\{\mathcal{C}\theta(y>x)\Big[
F^T\bigg(2+2\ln\frac{x}{y}+
\frac{1}{2}\ln^2\frac{x}{y}+2S\Big(\frac{x}{y}\Big)$
\\ && $\Ds
-\ln\Big(1-\frac{x}{y}\Big)
-\frac{1}{2}\ln^2\Big(1-\frac{x}{y}\Big)\bigg)$
\\
&& $\Ds
-\frac{1}{y}\ln^2\Big(1-\frac{x}{y}\Big)-\frac{2}{y}\ln\Big(1-\frac{x}{y}\Big)\Big]\Bigg\}_+
 $\\~
\end{tabular}

\\ \hline
\Large{r}\raisebox{-1.5cm}{\includegraphics[scale=0.33]{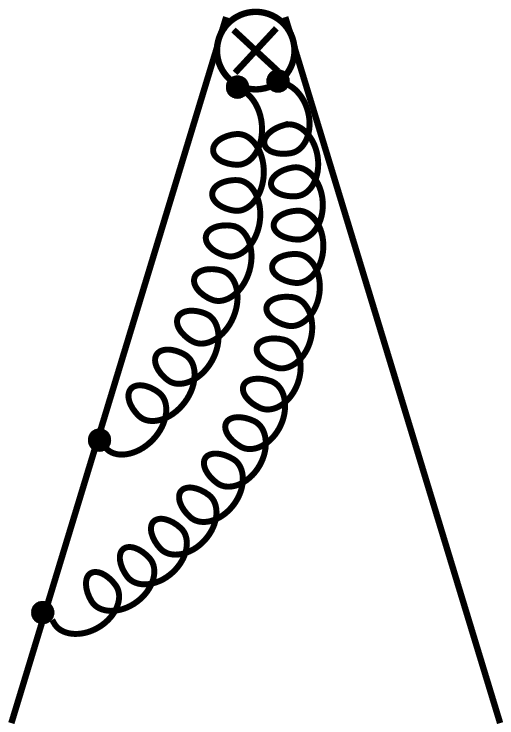}}

& \begin{tabular}{cc p{10cm}}
\\
 P(x) &=& $\Ds C_F\Big(C_F-\frac{C_A}{2}\Big)~\Big[p_0\Big(2\ln^2 x+4S(x)\Big)-16\ln\bar x\Big]_+
$\\&&$\Ds + C_FC_A\Big[p_0\Big(2+2S(\bar x)+\frac{1}{2}\ln^2
x+\ln\bar x+\frac{1}{2}\ln^2\bar x\Big)\Big]_+$

        \\\\
        \hline
        \\

 V(x,y)&=&$\Ds 2C_F\Big(C_F-\frac{C_A}{2}\Big)\cdot$
 \\&&
 $\Ds \Bigg\{\mathcal{C}\theta(y>x)\Big[F^T\bigg(2\ln^2\frac{x}{y}+4S\Big(\frac{x}{y}\Big)\bigg)
 -\frac{4}{y}\ln\Big(1-\frac{x}{y}\Big)\Big]\Bigg\}_+$
 \\&& $+C_FC_A\cdot$
 \\&& $\Ds \Bigg[\mathcal{C}\theta(y>x)F^T
 \bigg(4+4S\Big(1-\frac{x}{y}\Big)+\ln^2\frac{x}{y}+2\ln\Big(1-\frac{x}{y}\Big)$ \\ &&$\Ds+\ln^2\Big(1-\frac{x}{y}\Big)\bigg)\Bigg]_+
 $
\end{tabular}
\\ \hline
\Large{s}\raisebox{-1.5cm}{\includegraphics[scale=0.33]{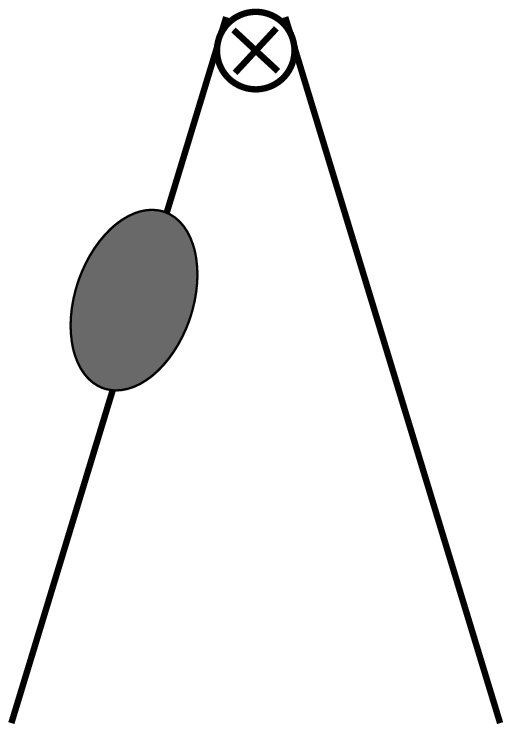}}

& \begin{tabular}{cc p{10cm}}
\\
 P(x) &=& $\Ds \Big(-\frac{17}{4}C_FC_A+\frac{3}{2}C_F^2+2C_FT_rN_f\Big)\delta(1-x)$

        \\\\
        \hline
        \\

 V(x,y)&=&$\Ds \Big(-\frac{17}{4}C_FC_A+\frac{3}{2}C_F^2+2C_FT_rN_f\Big)\delta(y-x)$
 \\~~
\end{tabular}
\\
\hline
 \end{longtable}
\vspace*{-8mm}
\begin{longtable}{||c|c||} ~~
~\begin{tabular}{c} \Large{a$^\star$}~~~~~~~
\Large{b$^\star$}~~~~~~~ \Large{c$^\star$}
~~~~~~~ \Large{j$^\star$} ~~~~~~~ \Large{p}\\
\raisebox{-1.5cm}{\includegraphics[scale=0.33]{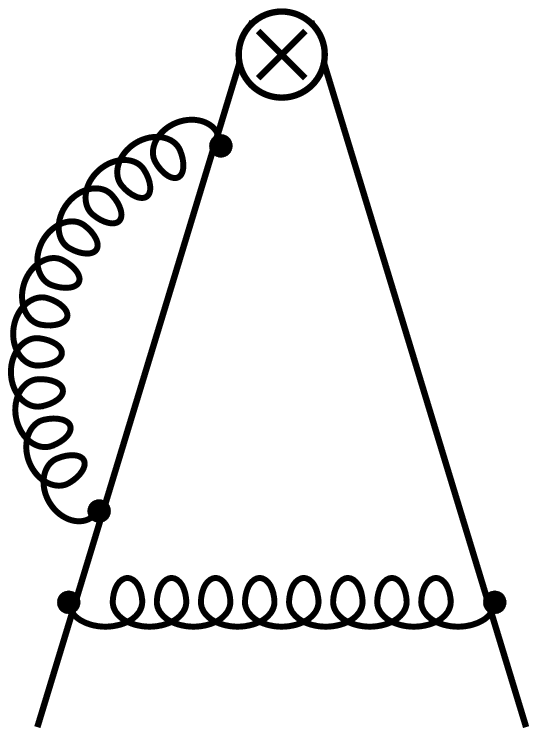}}
\raisebox{-1.5cm}{\includegraphics[scale=0.33]{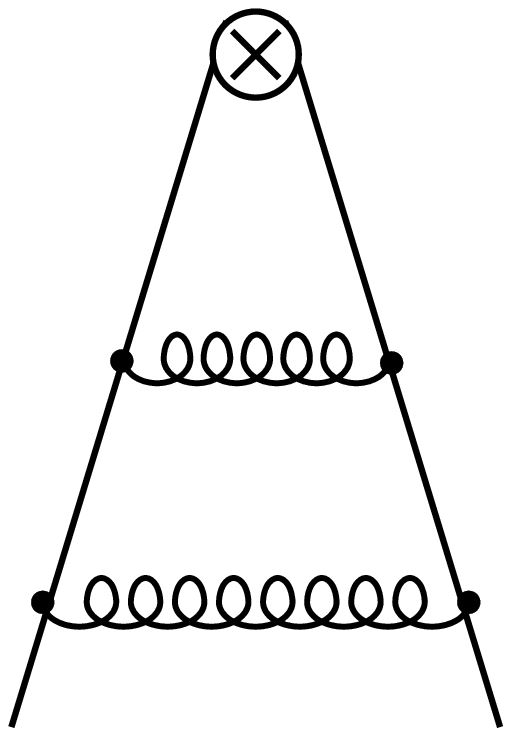}}
\raisebox{-1.5cm}{\includegraphics[scale=0.33]{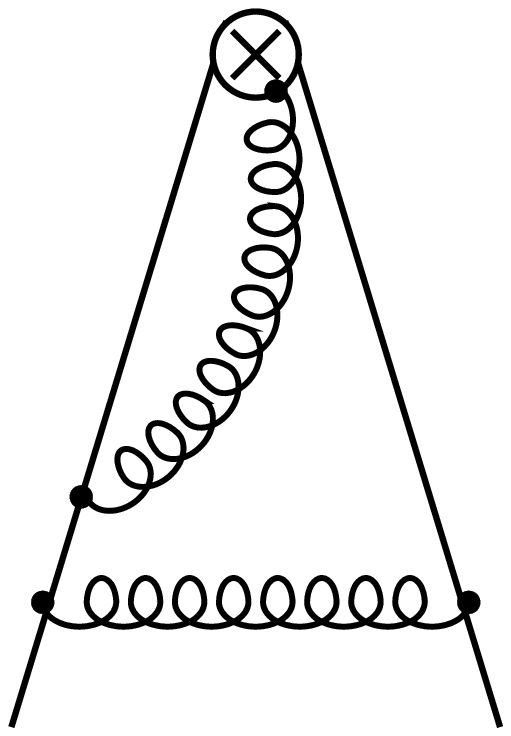}}
\raisebox{-1.5cm}{\includegraphics[scale=0.33]{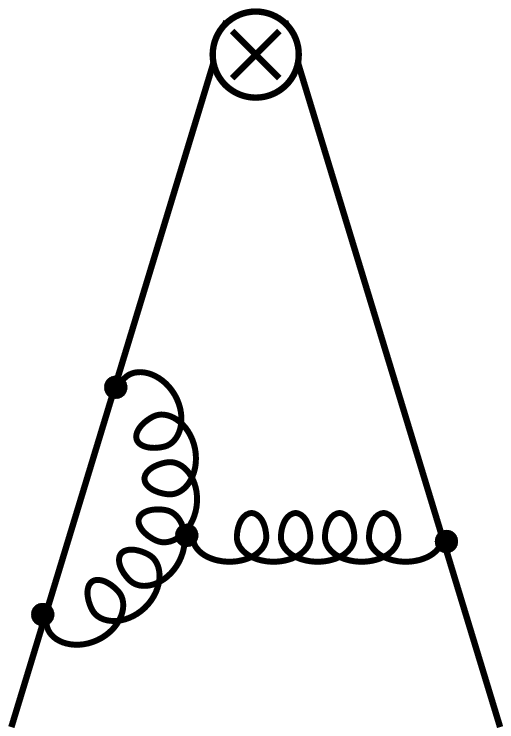}}
\raisebox{-1.5cm}{\includegraphics[scale=0.33]{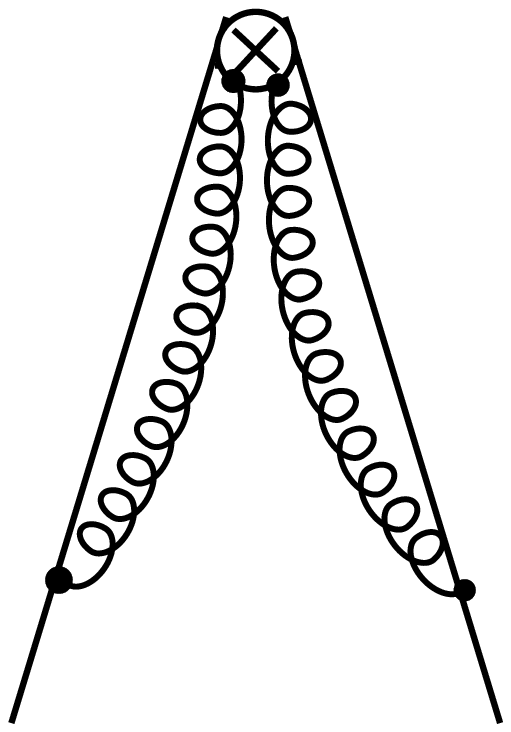}}
\end{tabular} & \begin{tabular}{cc p{2cm}}
\\
 P(x) &=& 0

        \\\\
        \hline
        \\

 V(x,y)&=& 0
 \\~~
\end{tabular}\\\hline
\end{longtable}
\noindent
Here $S(x) \equiv\text{Li}_2(x)-\text{Li}_2(1)$ and the special notation $G^T(x)$ will be clarified
in the next section (Eq.\ (\ref{eq:G2})).

There is, in general, a mixing of quark and antiquark densities in higher-loop calculations.
At NLO, the diagrams {\large{k}$^\star$,~\large{o}$^\star$} contribute to the kernel $P_{1 qq}$
expressing the probability to find a quark inside a quark (at $\eta =0$),
they also contribute to the kernel
$P_{1 q\bar{q}}$ giving the probability to find an antiquark inside a quark.
Actually, in the latter case, one should consider two kernels, viz.,
$P_{\pm}=P_{1 qq}\pm P_{1 q\bar{q}}$
for $\eta =\pm 1$, \cite{Vogelsang98}.
We shall separate these contributions and focus on the results for $P_{1 qq}$ and
$V^{T}_1$ in the next section.

\section{T\lowercase{he structure of the evolution kernels in} NLO}
\label{sec:structure}
In this section, we discuss the total results of the 
two-loop calculation and also the general structure
of the evolution kernels at NLO.
We commence with those elements that appear in the renormalization procedure 
at NLO of both the kernels $P$ and $V$.

\subsection{DGLAP kernel}
Collecting the ``quark-quark'' contributions to the NLO DGLAP kernel,  presented above, leads
for $P^{T}_{1 qq}$ to the final  expression
\begin{eqnarray}
\label{P1-res}
P^{T}_{1 qq}(x)&=&C_F^2 \cdot P^T_F(x) + C_FC_A \cdot P^T_G(x)+ C_FN_fT_r \cdot P^T_N(x),
\end{eqnarray}
where
\begin{eqnarray}
P^T_F(x)&=&~~4\bar{x}-\left[p_0(x)\Big(3 \ln(x)+4
\ln(x)\ln(\bar{x})\Big)\right]_+
+\delta(\bar{x}) \Big(\frac{43}{2}+8\zeta(3)-\frac{8\pi^2}{3}\Big)\ ,\label{P1F-res} \\
P^T_G(x)&=&-2\bar{x}+\! \left[p_0(x) \left(\ln^2(x)+\frac{11}{3}\ln(x)+
\frac{67}{9}-\frac{\pi^2}{3}\right)\right]_+ \!\!
-\delta(\bar{x}) \Big(\frac{365}{18}-\frac{4\pi^2}{3}+4\zeta(3)\Big),\label{P1G-res}\\
P^T_N(x)&=& -\frac{4}{3}\left[p_0(x)\left(\ln(x)+\frac{5}{3}\right)\right]_+
+\frac{26}{9}~\delta(\bar{x})\ . \label{P1N-res}
\end{eqnarray}
This expression together with the expression for $P_{1q\bar{q}}^T$ can be reduced,
after some simple algebraic manipulations, to those found in \cite{Vogelsang98}.

Let us rewrite the expression for $P^T_{1qq}$ in Eqs.\ (\ref{P1-res})--(\ref{P1N-res})
in such a form that corresponds to
the structure of $\hat{K}_1 R'$ at the two-loop level---see Eq.\ (\ref{Z}).
Following \cite{MR86ev}, we consider the
renormalization of the diagram $\Gamma$ and its contracted one-loop subgraph $E$
that can be written symbolically as $\Gamma=E \cdot W$,
where $W$ is the one-loop remainder.
As the result, the pole part of $E$
should be multiplied by the finite part of the remainder $W$
and vice versa.
All those subgraphs $E_i$
that are related to the charge renormalization of the intrinsic vertex in the various diagrams
(see, \eg, diagrams {\large d, g, h, l, m} in the list)
contribute to the coefficient of the QCD $\beta$-function $b_0=\frac{11}{3}C_A-\frac{4}{3}T_rN_f$.
After contracting each of these $E_i$ terms, the remainder reduces to one of the one--loop diagrams
{\large a, b, c} from Table \ref{tab:1-loop}.
The appropriate finite part of each of these diagrams in dimensional regularization can be obtained
from the differentiation of the auxiliary kernel (cf. similar kernels in \cite{MR86ev})
$P(x;\varepsilon)=4x^{1+\varepsilon} \bar{x}^{-1}$ with respect to the parameter
$\varepsilon$:
\begin{eqnarray}
\dot{p}_0(x)&=&\frac{d}{d\varepsilon}P(x;\varepsilon)\Big|_{\varepsilon=0} = p_0(x)\ln(x)\ .
\end{eqnarray}
On the other hand, the composite operator illustrated, \eg, in
diagrams {\large a$^\star$}, {\large b$^\star$}, {\large c$^\star$},
{\large o$^\star$}, {\large q}, {\large r} 
calls for a different sort of renormalization.
Notably, the contracted subgraph $E$ should include the composite operator
that coincides with that in the one-loop diagrams in Table\ \ref{tab:1-loop}.
The latter generates
the kernel $P^T_0$ (or $V^T_0$),
while the finite part of the remainder is formed {from the finite part of
$ \frac{1}{\varepsilon}P(x;\varepsilon)$, \ie, by $\dot{p}_0$
that finally leads to the contribution
\begin{eqnarray}
\left(\dot{p}_0 \ast (p_0)_+ \right)(x)&=&p_0(x)\left(4\ln(x)+4\ln(x)\ln(\bar{x})- 2\ln^2( x)\right).
\end{eqnarray}
Here the symbol $\ast$ denotes the Mellin convolution,
$\left(f\ast g \right)(x)=\int^1_0 dz dy \delta(x-y z)f(z)g(y)$.
Collecting all these terms together\textbf{,} one recasts $P^T_1$
in the form given by the first term in the curly brackets
below:
\begin{subequations}
\label{eq:P1-structure}
\begin{eqnarray}
P^T_{1qq}(x)&=&~\Bigg\{C_F\dot{p}_0 \ast \Big[b_0 \1  -
P^T_0\Big]+p_0(x)C_F\left[C_A\left(\frac{67}{9}-\frac{\pi^2}{3}\right)-\frac{20}{9}N_fT_r
\right]
\Bigg\}_+ \label{eq:P1-generalstructure} \\
&&
+ C_F \left(C_F - \frac{C_A}{2}\right)\left[4\bar{x}- 2\left(p_0(x)\ln^2( x)\right)_+\right]
\label{eq:P1-remainderstructure}\\
&& + \delta(\bar{x})C_F \left[C_F\frac{43}{2}-C_A\frac{365}{18}+ N_f
T_r\frac{26}{9} +\big(C_F-\frac{C_A}{2}\big)8\Big(\zeta(3)-\frac{\pi^2}{3}\Big)
\right]\ . \label{eq:P1-deltastructure}
\end{eqnarray}
\end{subequations}
The second term in the curly brackets in (\ref{eq:P1-generalstructure}) originates from the
product of the finite parts of the contracted subgraphs $E_i$,
or, more specifically, from the finite part of the charge renormalization
(diagrams {\large g}, {\large h}, {\large l}, {\large m})
and another finite and specific
(see diagrams {\large n$^\star$}, {\large o$^\star$}, {\large q}, {\large r})
part of the composite operator,
as well as from the pole parts of the remainder that are proportional to $p_0$.
In this respect, the coefficient of $p_0$ appears to be proportional
\cite{Korch89} to the two-loop cusp anomalous dimension \cite{KorchRad87},
\begin{eqnarray} \label{eq:gammacusp}
\frac{1}{4}\Gamma^{(1)}_{\rm cusp}=C_F\left[
C_A\ \left(\frac{67}{9}-\frac{\pi^2}{3}\right)-\frac{20}{9}N_fT_r\right].
\end{eqnarray}
The terms in (\ref{eq:P1-remainderstructure}) are formed by the diagrams
{\large k$^\star$} and {\large o$^\star$} with nonplanar elements
that also contribute to the ``quark-antiquark'' part $P^T_{1q\bar{q}}$ of the kernel.
Finally, the $\delta$-function in (\ref{eq:P1-deltastructure}) manifests the fact
that the corresponding local current is not conserved.
Let us emphasize at this point that the expressions in the r.h.s. 
of (\ref{eq:P1-generalstructure}), (\ref{eq:P1-deltastructure})
together with Eq.\ (\ref{eq:gammacusp})
has the general structure of any nonsinglet NLO DGLAP kernel that follows from
the renormalization procedure.
\subsection{ERBL kernel}
Collecting the partial contributions from the diagram-expansion list we arrive at
\begin{eqnarray}
V^T_1(x,y)&=&C_F^2 \cdot V^T_F(x,y) + C_FC_A \cdot V^T_G(x,y)+ C_FN_fT_r
\cdot V^T_N(x,y),
\end{eqnarray}
\begin{eqnarray}
V^T_F(x,y)&=&4\mathcal{C}\Big[\theta(y>x)\frac{x}{y}+\theta( y>\bar{x})\frac{\bar
x}{y}\Big] + 2 G^T(x,y)
+4\Bigg\{\mathcal{C}\theta(y>x)\Big[F^{T}\ln^2\frac{x}{y}\nonumber \\
&& +\frac{1}{y\bar y}\ln x\ln \bar x - \frac{3}{2}F^{T}\ln\frac{x}{y}
-\big(F^{T}-\bar F^{T}\big)\ln\frac{x}{y}\ln\Big(1-\frac{x}{y}\Big)\Big]\Bigg\}_+ \nonumber \\
&& +4\Big[\frac{11}{8}+6\zeta(3)-2\frac{\pi^2}{3}\Big]\delta(x-y)\ ,
\end{eqnarray}

\begin{eqnarray}
V^T_G(,x,y)&=& - 2 \mathcal{C}\Big[\theta(y>x)\frac{x}{y}+\theta( y>\bar{x})\frac{\bar
x}{y}\Big]- G^T(x,y) \nonumber \\
& &+ \Big[\mathcal{C}\theta(y>x)2F^{T}\Big(\frac{11}{3}
\ln\frac{x}{y}+\frac{67}{9}-\frac{\pi^2}{3}\Big) \Big]_+ \nonumber  \\
&&  +\Big[-\frac{221}{18}-12\zeta(3)+4\frac{\pi^2}{3}\Big]\delta(y-x)\ ,  \\
V^T_N(x,y)&=&-\frac{4}{3}\left[\mathcal{C}\theta(y>x)2F^{T}\Big(\ln\frac{x}{y}+\frac{5}{3}\Big)\right]_+
+\frac{26}{9}\delta(y-x) \ .
\end{eqnarray}
Here and below
\begin{eqnarray}
G^T(x,y)&=&-4\mathcal{C}\Big[\theta(y>x)\Big(\bar F^{T} \ln \bar
x\ln y-F^{T}\big[\text{Li}_2(x)+\text{Li}_2(\bar
y)\big]+\frac{\pi^2}{6}F^{T}\Big)+\theta(\bar y < x)~~~~~~~~~~~~~~~~~ \nonumber  \\
&&~\times\Big((F^{T}-\bar
F^{T})\big[\text{Li}_2(1-\frac{x}{y})+\frac{1}{2}
\ln^2x\big]+F^{T}\big[\text{Li}_2(\bar y)-\ln x\ln y\big]+\bar
F^{T}\text{Li}_2(\bar x)\Big) \Big] . \label{eq:G2}
\end{eqnarray}
The term $G^T$ is ``diagonal'',
\ie, $y \bar{y}G^T(x,y) = x \bar{x} G^T(y,x)$,
$G^T$ coincides with the similar term $G$ in
the unpolarized kernel $V_1$ \cite{MR85} by performing there \cite{BFM2000} the replacement $F^T \to F$
and excluding the third term $\frac{\pi^2}{6}F^{T}$ in the first line of (\ref{eq:G2}).
This term is tied to $G^T$ in order to preserve} $\Gamma_{\rm cusp}$
in the general structure of $V_1^T$.

The origin of the structure of the ERBL kernel can be considered by analogy with the DGLAP case,
as explained in \cite{MR86ev}.
Taking into account that the ERBL auxiliary kernel is
$V(x,y;\varepsilon)= 1/2\mathcal{C}\left(\theta(y>x)P(x/y;\varepsilon)/y \right)$
leads to the following finite part of the corresponding remainder
$\dot{V}^{T}_0=1/2\mathcal{C}\left(\theta(y>x)p_0(x/y)/y\right)$,
which, after introducing an appropriate convolution for the one-loop elements \cite{MR86ev},
$\left(f\otimes g \right)(x,y)=\int^1_0 dz f(x,z)g(z,y)$, leads to the expression
\begin{subequations}
\label{eq:V1-structure}
\begin{eqnarray}
V^T_1(x,y)&=&\Ds \bigg\{C_F \dot{V}^{T}_0\otimes\Big( b_0 \1 -V^{T}_{0}\Big)
+\mathcal{C}\theta(y>x)2F^{T}\ \frac{\Gamma^{(1)}_{\rm cusp}}{4}
+C_F [g_+,\ \otimes V^{T}_{0}]\bigg\}_+ \label{eq:V1-generalstructure}  \\
&&+~C_F\Big(C_F-\frac{C_A}{2}\Big)
\left\{4\mathcal{C}\Big[\theta(y>x)\frac{x}{y}+ \theta( y>\bar{x})\frac{\bar
x}{y}\Big] + 2 G^T(x,y)\right\}  \label{eq:V1-remainderstructure}\\
&&+~\delta(y-x)C_F \left[C_F\frac{27}{2}-C_A\frac{221}{18}+ N_f
T_r\frac{26}{9} \right]\ . \label{eq:V1-deltastructure}
\end{eqnarray}
\end{subequations}
The structure of the elements of $V^T_1$ in (\ref{eq:V1-generalstructure})-(\ref{eq:V1-deltastructure})
 resembles that of $P^T_1$ in (\ref{eq:P1-generalstructure})-(\ref{eq:P1-deltastructure})
with a natural replacement
of notation for the convolution and the symbols
$\otimes \to \ast,~\dot{V}^{T}_0 \to \dot{p}_0,~V^{T}_{0} \to P^{T}_{0},~2F^{T} \to p_0$
with the exception of the important third term in the curly bracket in (\ref{eq:V1-generalstructure}).
In addition to the ``nondiagonal'' term proportional to $b_0$
and to the proper operator renormalization (see the first convolution in
Eq.\ (\ref{eq:V1-generalstructure})),
there appears an additional ``nondiagonal'' term which is represented by the commutator
$[g_+,\ \otimes V^{T}_{0}] \equiv g_+ \otimes V^{T}_{0} - V^{T}_{0} \otimes g_+$.
This term is induced by the leading-order anomaly in 
special conformal transformations of conformal operators,
\begin{eqnarray}
g(x,y)&\!\!\!=\!\!\!&  - 2\mathcal{C}\frac{ \theta(y>x)}{y-x}
\ln\left(1-\frac{x}{y}\right)\ ,
\end{eqnarray}
an interesting issue explained in \cite{BFM2000,Mul95}.
All the other terms in $V^T_1$ are ``diagonal''.
Concluding these considerations let us mention that the expressions in
the r.h.s. of (\ref{eq:V1-generalstructure}), (\ref{eq:V1-deltastructure})
give us the elements of the general structure of any NLO ERBL kernel.

\section{E\lowercase{ffects of two-loop evolution for the meson} DA }
 \label{sec:evolution}
The subject of this section concerns the effects of the two-loop QCD evolution in
an appropriate example inspired by calculations
of the $B \to \rho~\nu e$ decay \cite{BM2000,BB97}.
For the leading-twist DA of the transversely polarized
$\rho$--meson
expanded in a Gegenbauer series
\begin{equation}
\varphi(x,\mu^2_0)~=~\sum_{n=0} c_n(\mu^2_0)\psi_n(x)
\end{equation}
the two-loop evolution of each harmonic $\psi_n$ from $\mu^2_0$ to $\mu^2$,
 $\psi_n(x) \to \Phi_n(x,\mu^2)$, can be approximately represented%
 \footnote{The NLO evolution that preserves the renormalization-group property
 for the ``nondiagonal'' elements was worked out in \cite{BS2005}.}
as
\cite{MR86ev}
\begin{equation}\label{evolut2loop}
\Phi_n(x,\mu^2)~=~\exp\Bigg(-\int_{a_s(\mu_0^2)}^{a_s(\mu^2)}d\alpha \
\frac{\gamma(n,\alpha) }{\beta(\alpha)}\Bigg)
\left[\psi_n(x)+
a_s\sum_{m>n}\frac{d_{mn}}{N_m}\psi_m(x)\right]~.
\end{equation}
Here we have
\begin{eqnarray}
\gamma(n,a_s)~&=&~a_s\ \gamma^T_{0}(n)+a_s^2\ \gamma^T_{1}(n)~, \\
\beta(a_s)~&=&~-a_s^2 \ b_0-a_s^3 \ b_1~,
\end{eqnarray}
$$d_{mn}~=~\frac{Z_{mn}}{\gamma_{0}(n)-\gamma_{0}(m)-b_0}
\Bigg[1-\Bigg(\frac{a_s(\mu^2)}{a_s(\mu_0^2)}\Bigg)^{(\gamma_{0}(n)-\gamma_{0}(m)-b_0)/b_0}\Bigg]~,$$
$$Z_{nm}~=~C_n^{3/2}\otimes V^T_{1} \otimes \psi_m~,
~~N_n\,\delta_{n m}~=~C_n^{3/2} \otimes \psi_m~=~\frac{(n+1)(n+2)}{4(2n+3)}\delta_{n m}~$$
and $\gamma(n)$ is the anomalous dimension with
$\Ds \gamma^T_1(n)~=~Z_{nn}/N_n$.
The coefficients $d_{nm}/N_m$ can be calculated
analytically (in the form of lengthy sums) by virtue of
the knowledge of the structure of the ``nondiagonal'' and ``diagonal'' terms
in expressions
(\ref{eq:V1-generalstructure})-(\ref{eq:V1-remainderstructure}).
Their evaluation for the values of the input parameters $\mu_0^2=1$\ GeV$^2$,
$\mu^2_{B}=36$\ GeV$^2$ (the latter being the characteristic scale of the  $B \to \rho~\nu e$ decay)
for $N_f~=4$ is presented in Table 2.

\begin{tabular}{cc}
$
\begin{array}{l|cccc}
m\setminus n & 0 & 2 & 4& 6 \\
\hline
2 & -0.398 & 0 & 0 & 0 \\
4 &  -0.013 & -1.08 & 0 & 0 \\
6 & 0.024 & -0.297 & -1.269 & 0 \\
8 & 0.024 & -0.094 & -0.485 & -1.288 \\
10 & 0.02 & -0.026 & -0.216 & -0.585 \\
12 & 0.015 & -0.001 & -0.103 & -0.303 \\
14 & 0.012 & 0.008 & -0.049 & -0.168 \\
16 & 0.01 & 0.011 & -0.022 & -0.097 \\
18 & 0.008 & 0.012 & -0.007 & -0.057 \\
20 & 0.006 & 0.012 & 0.000 & -0.033
\end{array}
$ &\vspace{5pt}

\begin{picture}(100,100)
\put(-60,-60){\includegraphics[scale=0.47]{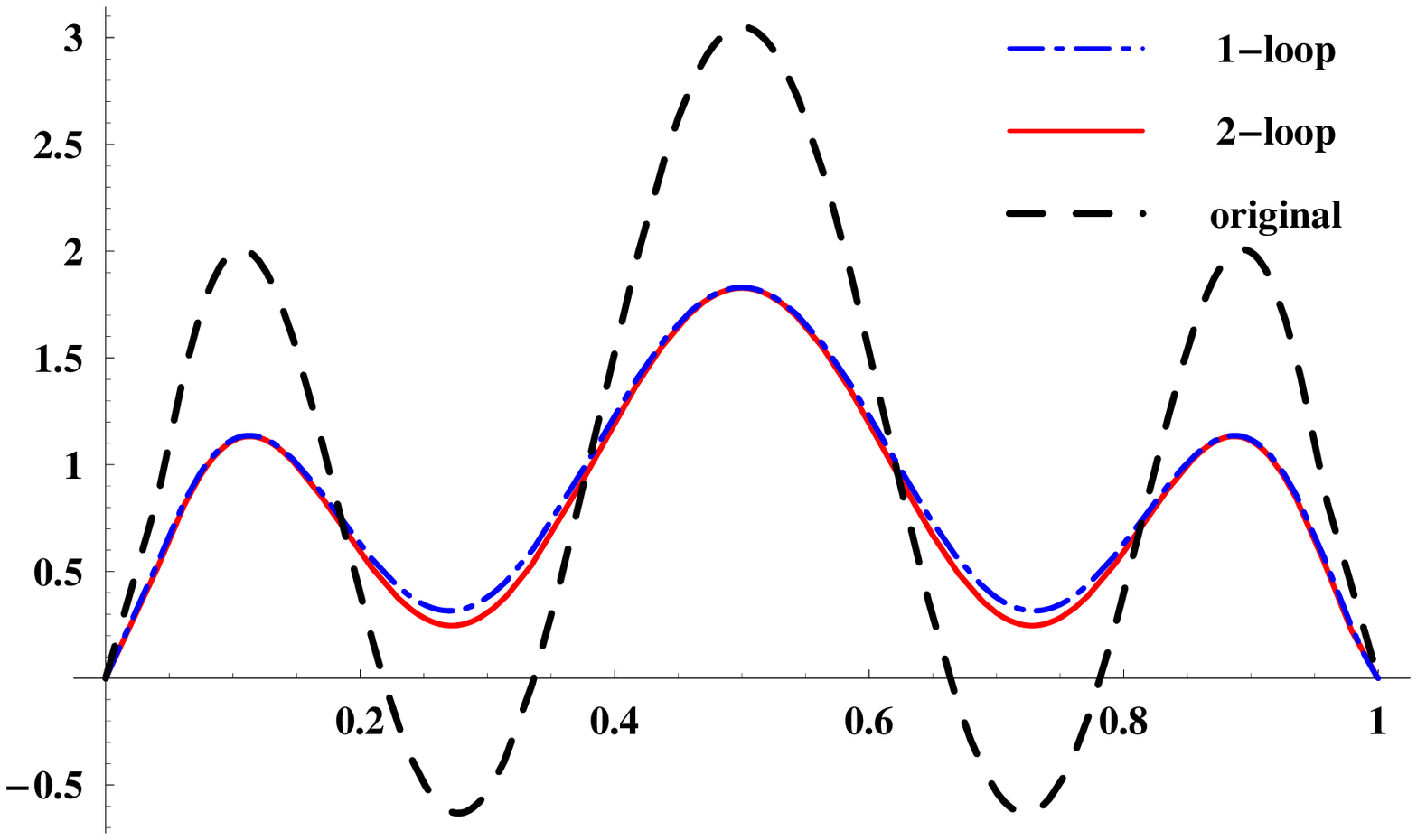}}
\put(-10,70){\makebox{$\varphi(x)$}} \put(65,-45){\makebox{$x$}}
\end{picture} \\
Table 2. \footnotesize{Numerical values of the $d_{nm}/N_m$ coefficients}&~Fig. 1
\footnotesize{The result of one-loop and two-loop evolution} \\
           \footnotesize{for the first 6 harmonics $\psi_n$.}&
           \footnotesize{ of $\varphi^{T}_{\rho}(x;\mu^2_{0}=1\, \text{GeV}^2)$ to
           $\mu^2_{B}=36$\, GeV$^2$ }
\end{tabular}
\vspace{2pt}

\noindent These coefficients $d_{nm}/N_n$ decrease not so fast as those for the
unpolarized case \cite{MR86ev}.
The numerical calculation shows that
truncating the sum in (\ref{evolut2loop}) after the 10th term provides
us with a $0.03\%$ accuracy at the scale $36$~GeV$^2$.

To get an estimate for the difference between the two-loop and the one-loop result, let us compare
the corresponding first inverse moments of the DA, $\langle x^{-1}\rangle=\int_0^1
\varphi(x)/x dx$.
The ratios $(\langle
x^{-1}_{2-{\rm loop}}\rangle/\langle x^{-1}_{1-{\rm loop}}\rangle-1)\%$ at this
scale are $-4.1\%$ for $\psi_0(x)$, $-1.4\%$ for
$\psi_2(x)$, $-0.3\%$ for $\psi_4(x)$, and even less for higher
harmonics.
As regards the model distribution amplitude
$\varphi^{T}_{\rho}$ normalized at $\mu^2_{0}\simeq 1$\ GeV$^2$, 
$$\varphi^{T}_{\rho}(x;\mu^2_{0})=\psi_0(x)+0.29\psi_2(x)+0.41\psi_4(x)-0.32\psi_6(x)\ ,$$
which was obtained obtained
for a transversely polarized $\rho$--meson in \cite{BM2000},
the ratio of the first inverse moments takes the value $3.6\%$.
The evolution effect on the meson distribution amplitude $\varphi^{T}_{\rho}(x;\mu^2_{0})$
is illustrated in Fig.\ 1.
The dashed black line shows the unevolved expression, while the result of the
two-loop evolution to the scale $\mu^2_{B}$ is represented by a solid red line,
and the one-loop result is shown as a blue dashed-dotted line.

\section{C\lowercase{onclusions}}
\label{sec:conclusion}
Let us summarize our findings.
In Sec.\ \ref{sec:diagrams}, 
we presented the diagram-by-diagram results of a direct two-loop calculation of
the DGLAP, $P^T$, and the ERBL, $V^T$, evolution kernels
for transversity distributions, employing the Feynman gauge.
The mutual correspondence between the $V$ and $P$
results, for each of the diagrams, was checked making use of
the relation
$
P(z)
=
 \lim_{\eta\to 0} \frac{1}{|\eta|}
V
\left( \frac{z}{\eta}, \frac{1}{\eta} \right)
$, \cite{MRGDH1998}.
It was found that the total result for $P_1^T$ coincided with the one in
\cite{Vogelsang98} (obtained within a light-cone gauge calculation), whereas the 
total result for $V_1^T$ turned out to agree
with the prediction obtained in \cite{BFM2000}.

We worked out the general structure of any nonsinglet NLO DGLAP kernel,
Eqs.\ (\ref{eq:P1-structure}),(\ref{eq:gammacusp}),
and any NLO ERBL kernel, Eqs.\ (\ref{eq:V1-structure}), respectively,
subject to the renormalization procedure.

The NLO evolution of the DA of twist 2 (for transversely polarized $\rho$-meson)
was considered and its relative effect was estimated for the 
inverse moment of the corresponding DA.
This effect  amounts to a few per cents (4\% for the zero
Gegenbauer harmonic) after evolving from the low scale
$\mu^2_{0}\simeq1$~GeV$^2$
to the characteristic scale $\mu^2_{B}=36$~GeV$^2$ of the B-decay process.

\noindent \textbf{Acknowledgments.}
We would like to thank D. M\"uller for
stimulating discussions and useful remarks,
 A. Grozin and N. G. Stefanis 
for the careful reading
of the manuscript and clarifying criticism.
We are indebted to Prof.\ Klaus Goeke
for the warm hospitality at Bochum University, where this work was
partially carried out.
This work was supported in part by RFBR
(grants No.\ 06-02-16215, No.\ 07-02-91557 and No.\ 08-01-00686),
the Heisenberg--Landau Programme (grants 2006--2008),
the COSY Forschungsprojekt J\"ulich/Bochum, and the Deutsche
Forschungsgemeinschaft (Project DFG 436 RUS 113/881/0-1).




\end{document}